\documentclass[apj]{emulateapj}
\usepackage{hyperref}
\newcommand{\rosFormatApJ}[1]{}
\newcommand{\rosFormatGeneral}[1]{#1}

\newcommand{\rosFormatOptions}[2]{\rosFormatGeneral{#1}\rosFormatApJ{#2}}

\begin{document}

\newcommand\abbrvPSconstraints{PSC1}
\newcommand\abbrvPSutility{PSF1}
\newcommand\abbrvPSutiltwo{PSF2}
\newcommand\abbrvChungleeNSNSa{KKL}
\newcommand\abbrvStarTrack{BKB}

\newcommand {\peryr}{$\rm{yr}^{-1}$}
\newcommand {\permyr}{$\rm{Myr}^{-1}$}
\newcommand{\nsns}{NS$-$NS\ }
\newcommand{\nswd}{NS$-$WD\ }
\newcommand{\wdns}{WD$-$NS\ }
\newcommand{\tlife}{\tau_{\rm life}}
\newcommand{\tmrg}{$\tau_{\rm mrg}$}
\newcommand{\td}{$\tau_{\rm d}$}
\newcommand{\tc}{$\tau_{\rm c}$}
\newcommand{\tsd}{$\tau_{\rm sd}$}
\newcommand{\rate}{$\cal R$}
\newcommand{\rtot}{${\cal R}_{\rm tot}$}
\newcommand{\rpeak}{${\cal R}_{\rm peak}$}
\newcommand{\rdet}{${\cal R}_{\rm det}$}
\newcommand{\ntot}{$N_{\rm tot}$}
\newcommand{\npsr}{N_{\rm PSR}}
\newcommand{\nobs}{$N_{\rm obs}$}
\newcommand{\nmean}{$<N_{\rm obs}>$}
\newcommand{\solarM}{M$_{\rm \odot}$}
\newcommand{\chmass}{$\cal M$}
\newcommand{\fmin}{$f_{\rm min}$}
\newcommand{\fmax}{$f_{\rm max}$}
\newcommand\unit[1]{\, {\rm #1}}

\title{Constraining population synthesis models via
  empirical binary compact object merger and  supernovae  rates}
\date{7/30/2007}
\author{R.\ O'Shaughnessy}
\affil{Northwestern University, Department of Physics and Astronomy,
  2145 Sheridan Road, Evanston, IL 60208, USA}
\email{oshaughn@northwestern.edu}
\author{C. Kim}
\affil{Cornell University, Department of Astronomy, 522
  Space Sciences Building, Ithaca, NY 14853, USA}
\author{V. Kalogera}
\affil{Northwestern University, Department of Physics and Astronomy,
   2145 Sheridan Road, Evanston, IL 60208, USA}
\and
\author{K. Belczynski}
\affil{Tombaugh Fellow, New Mexico State University, Las
   Cruces, New Mexico, 88003, USA}

\begin{abstract}
The observed samples of supernovae (SN) and double compact objects (DCOs)
provide several critical constraints on  population-synthesis
models: the parameters of these models must be carefully chosen to
reproduce, among other factors, (i) the formation rates of double neutron
star (NS-NS) binaries and of white dwarf-neutron star (WD-NS)
binaries, estimated from binary samples, and  (ii) the type II and Ib/c supernova rates.
Even allowing for extremely
conservative accounting  of the
uncertainties in observational and theoretical predictions,
we find only a few plausible population synthesis models  
 (roughly 9\%) are consistent  with 
DCO and SN rates empirically determined from  
observations.
As a proof of concept, we describe the information that
can be extracted about population synthesis models given these 
observational tests, including surprisingly good agreement with the
neutron star kick distributions inferred from pulsar proper-motion measurements. 
In the present study, we
find that the current observational constraints favor: kicks described
by a single Maxwellian with a characteristic velocity of about
350km/s (i.e., at maximum likelihood; kick velocities between 100km/s
and 700km/s remain within the 90\% confidence interval of unimodal distributions);
mass-loss fractions during non-conservative, but stable, mass transfer
episodes of about 90\%; and common envelope parameters of about
 0.15-0.5. 
Finally, we use the subset of astrophysically consistent models to
predict the rates at which  black
hole-neutron star (BH-NS) and NS-NS binaries  merge in the Milky Way
and the nearby Universe, assuming Milky-Way-like galaxies dominate.
Inevitably, the resulting probability distributions for merger rates
depend on our assumed priors for the population model input
parameters. In this study we adopt relatively conservative priors
(flat) for all model parameters covering a rather wide range of
values. However, as we gain confidence
in our knowledge of these inputs,  the range of merger rates
consistent with our knowledge should shift and narrow.
\end{abstract}

\maketitle

\section{Introduction}
Many ground-based gravitational wave detectors now operating at
or near design sensitivity (i.e., LIGO, GEO, TAMA, and VIRGO) are 
designed to detect the late stages of double compact object (DCO) 
inspiral and final merger.  Based only on early-stage data, these
instruments have already provided conservative upper limits to certain
DCO merger rates in the Milky Way
\citep[see,e.g.][]{LIGOS2macho,LIGOS2nsns}.  
If LIGO indeed maintains its current level of sensitivity
(Marx, for the LIGO Scientific Collaboration, 2006),
LIGO data could 
 set rate limits that exclude optimistic theoretical predictions, such
 as a population of binary $\approx 10 M_\odot$ black
   holes merging at a present-day rate density above $\simeq 1 {\rm
     Mpc}^{-3} {\rm Myr}^{-1}$ 
(based
 on the design LIGO range of $15 {\rm Mpc}$ for double neutron
 stars). 
This upper limit  has occasionally been reached in some theoretical
predictions for BH-BH merger rates from isolated
\citep[e.g.][hereafter \abbrvStarTrack]{StarTrack}
and interacting \citep[e.g.][]{clusters-2005} binary stellar systems.%
\footnote{More generally,  
\citet{PSutility,PSconstraints,1995ApJ...440..270B,1998AA...332..173P}; \cite{Chris-nspaper-2007}; and references
therein discuss the range of  DCO
merger rates expected to arise from binary evolution in Milky Way-like
galaxies.}
Thus, gravitational-wave based upper limits
(and, eventually, detections)  are expected to provide constraints on
theoretical models of DCO formation.

Lacking a sample of  DCOs containing a black hole,
the only route to BH-BH merger rates is via population synthesis models.
These involve a Monte Carlo exploration of the likely
life histories of binary stars, given 
statistics governing the initial conditions for binaries and 
a method for following the behavior of single and
binary stars 
\citep[see, e.g., \abbrvStarTrack,][and references therein]{2002MNRAS.329..897H,1996AA...309..179P,1998ApJ...496..333F}.
Unfortunately, our understanding of the evolution of single and binary
stars is incomplete, and we parameterize the associated uncertainties with a great
many parameters ($\sim 30$), many of which can cause the predicted
DCO merger rates to vary by more than one or two orders of
magnitude when varied independently through their plausible range.
To arrive at more definitive answers for DCO merger rates,
we must substantially reduce our uncertainty in the parameters that
enter into population synthesis calculations through comparison with
observations when possible. 

The simplest and most direct way to constrain the parameters
of a given population synthesis code is to compare several of  its
many predictions against observations.
For example, 
\citet{PSconstraints} (henceforth \abbrvPSconstraints)
required their population synthesis
models be consistent with the  empirically
estimated formation rates derived from  
 the three (now four) known Galactic \nsns binaries  which are tight
enough to merge through the emission of gravitational waves within
$10$\,Gyr.  
In this paper, we require our population synthesis models to be
consistent (modulo experimental error) with six observationally
determined rates: (i) the formation rate 
implied by the  known Galactic merging \nsns binaries, mentioned
above; (ii) the formation rate implied by the known Galactic
\nsns binaries which do \emph{not} merge within $10$ Gyr (henceforth denoted ``wide''
\nsns binaries); (iii,iv) the formation rate  implied by the sample of
merging and eccentric WD-NS binaries; and finally (v,vi) the type II and
type Ib/c SNe rates.
Further, we use the set of models consistent with these constraints to
revise our population-synthesis-based expectations for various DCO
merger rates, assuming no prior information, so all tested
population synthesis model parameters consistent with our constraints
are treated equally.
%
Though many of the core concepts have been presented
  previously, notably in 
\abbrvPSconstraints,
this paper presents  more constraints,
higher-quality fits, and  particularly the first  analysis of fitting
errors \citep[as developed in our companion paper][henceforth \abbrvPSutiltwo]{PSutil2} to
arrive at substantially more reliable rate predictions that satisfy
all available empirical rate constraints.  Additionally, we present both the one and two-dimensional
distributions of those population synthesis parameters which reproduce
our complete set of constraints.  

We note however, that we focus only on compact object binaries with
one or two neutron stars.  Merger rate predictions for binaries with
two black holes must account for the mix of both spiral (Milky-Way
like) and elliptical galaxies in the local Universe,
since present-day merging double black hole binaries are often substantially
longer-lived than their lighter counterparts; thus, compared to the
small fraction of extremely long-lived progenitors of merging low-mass  binaries,  a
significant fraction of presently merging BH-BH binaries  could have been born
 $\simeq 10\unit{Gyr}$ ago during the epoch of elliptical galaxy
formation.
As modeling of the latter part involves computational
elements and methodology that is not part of the present study (i.e.,
a two-component, time-dependent star formation model), we
intend to undertake BH-BH rate predictions in a forthcoming paper
\cite{PSellipticals}; see also \cite{ADM:de2005}.   
However, the evolutionary channels and essential physics are very
thoroughly addressed in \cite{ChrisBH2007}.
In Sections \ref{sec:data:DCOs} and \ref{sec:data:other} 
we describe the observational constraints we impose on the
population synthesis models: agreement with empirical
DCO merger rates [\nsns, based on \citet{Chunglee-nsns-1} 
(hereafter KKL) and 
\abbrvPSconstraints;
NS-WD, drawn
from \citet{Chunglee-wdns-proceedings,Chunglee-wdns-1}];  and
supernova rates [drawn from \citet{Cappellaro:SNa}].  
Considering the
significant one-sigma systematic error in the pulsar birthrate \citep{2004ApJ...617L.139V} and the strong
correlations of that observable to the already included DCO merger
rate observations, we do not  impose the pulsar birthrate as an
additional constraint.
In \S\ref{sec:ps} we briefly describe the version  of the
  StarTrack code used and
summarize the updated results from 
\abbrvPSutiltwo{}  
on our population synthesis models
and the resulting predictions for DCO formation rates and SN rates.
In Section \ref{sec:constraints:DCO} we  select from our family of possible
models those predictions which are consistent with observations of
DCOs.  We thoroughly review the information gained from
these constraints and discuss their implications for our understanding of binary evolution.
Also, we  employ this sample to generate
refined predictions for the  BH-NS and \nsns merger rates
through the emission of gravitational waves.

\section{Observations of NS DCOs}
\label{sec:data:DCOs}
Seven  \nsns binaries and four WD-NS binaries (with relatively massive WDs) have been
discovered so far in the galactic disk, using pulsar surveys with
well-understood selection effects. We are very specifically
\emph{not} including the recently-discovered binary PSR J1906+0746
\citep{psr:discovery:1907+0746}, because the companion cannot be
definitively classified as a WD or NS at present\footnote{%
Our predictions for the galactic NS-NS merger rate would increase by a
factor of roughly two if indeed PSR J1906+0746 were a double neutron star.%
}. We also omit
PSR J2127+11C found in the globular cluster M15, since its formation
is thought to be dynamical and not due to isolated binary evolution in
the Galactic field; \cite[see,e,g.][]{1990Natur.346...42A}.
\nocite{lrr-2005-7}
The basic measured and derived properties of these
binaries are presented in Table~\ref{tab:dcos}.

\rosFormatGeneral{\begin{deluxetable*}{lrrrrrrr|rrrrrrrc}[ht]
\centering}
\rosFormatApJ{\begin{deluxetable}{lrrrrrrrrrrrrrrc}}
\tablecolumns{16}
\tablewidth{300pc}
\tablecaption{Properties of compact binaries}
\tabletypesize{\footnotesize}
\tablehead{
\colhead{PSRs$^\tablenotemark{a}$} &
\colhead{$P_{\rm s}^{b}$} &
\colhead{{\it \.{P}}$_{\rm s}^{c}$} &
\colhead{$P_{\rm b}^{d}$ }&
\colhead{M$_{\rm PSR}^{e}$} &
\colhead{M$_{\rm c}^{f}$} &
\colhead{e$^{g}$} &
\colhead{$f_b^{h}$} &
\colhead{$\tau_{\rm c}^{i}$} &
\colhead{$\tau_{\rm sd}^{j}$} &
\colhead{$\tau_{\rm mrg}^{k}$} &
\colhead{$\tau_{\rm d}^{l}$} &
\colhead{N$_{\rm PSR}^{m}$} &
\colhead{$A^{n}$} &
\colhead{Refs$^{o}$}
 \\ 
\colhead{} &  \colhead{(ms)} &  \colhead{($10^{-18}$)} &  \colhead{(hr)} & \colhead{(M$_{\rm \odot}$)} & \colhead{(M$_{\rm \odot}$)} &
\colhead{}  & \colhead{}   
&  \colhead{(Gyr)} & \colhead{(Gyr)} &  \colhead{(Gyr)} & \colhead{(Gyr)} &  \colhead{}  &\colhead{(Myr)} &
\colhead{} 
}
\startdata
(1) NS-NS(m) & &&&&&&   &&&&&&\\
\hline
B1913+16   & 59.03  & 8.63 & 7.752   & 1.4408 & 1.3873  & 0.617 & 5.72    & 0.11  & 0.065 & 0.3   & 4.34 & 620  & 0.103 &   1,2\\
B1534+12   & 37.90  & 2.43 & 10.098  & 1.3332 & 1.3452  & 0.274 & 6.04    & 0.25  & 0.19  & 2.7   & 9.55 & 440  & 1.011 &    3,4,5\\
J0737-3039A & 22.70  & 1.74 & 2.454   & 1.337 & 1.25     & 0.088 & $6^*$   & 0.207 & 0.145 & 0.085 & 14.1 & 1620  & 0.023&   6,7\\
J1756-2251 & 28.46  & 1.02 & 7.67    & 1.4   & 1.18     & 0.181 & $6^*$   & 0.44  &  0.38 & 1.7  & 16  & -   & -  & 8 \\
\hline
(2) NS-NS(vw)& &&&&&&\\
\hline
J1811-1736 & 104.182 & 0.916       & 451.20  & 1.62 & 1.11 & 0.828  &$6^*$&1.8 & 1.74 & n/a  & 7.8 & 606   & 2.61 & 9,10 \\
J1518+4904 & 40.935  & 0.02        & 207.216 & 1.56 & 1.05 & 0.25   &$6^*$&32 & 32 & n/a & 54.2 & 282  &37.8  &  11 \\
J1829+2456 & 41.0098 & $\sim$ 0.05 & 28.0    & 1.14 & 1.36 & 0.139  &$6^*$&13.0 & 12.9 & n/a & 43.3 & 272  & 32.7 & 12 \\
\hline
(3) WD-NS(m) & &&&&&&\\
\hline
J0751+1807 &  3.479 & 0.00779&  6.312  & 2.1 &  0.191& $\approx 0$ &$6^*$ & 7.08 &  6.9 & 4.3 & 400 & 2850   & 0.66  &  13,14 \\
J1757-5322 &  8.870 & 0.0263&  10.879 & 1.35 & 0.67   & $\approx 0$ &$6^*$& 5.06 & 4.93 & 8.0 & 145 & 1240   & 1.74  &  15 \\
J1141-6545 &  393.9 & 4290  &  4.744  & 1.30  & 0.986  &  0.172     &$6^*$& 0.0015 & - & 0.56 & 0.104 & 370  & 0.047  &16,17 \\
\hline
(4) WD-NS(e)  & &&&&&&\\
\hline
J1141-6545  & 393.9    & 4290  &  4.744  &  1.30 & 0.986  &  0.172  &$6^*$& 0.0015 & - & 0.56 & 0.104 & 370    & 0.047 &  15,16 \\
B2303+46    &  1066.37 & 569   &  295.2  &  1.34 & 1.3   &  0.658   &$6^*$& 0.0297 & 0.014 & n/a & 0.138 & 240  & 0.106& 17,18\\
\enddata
\label{tab:dcos}
\tablenotetext{\rosFormatOptions{1}{a}}{Name of double-compact binary  observed via a
  component pulsar.}
\tablenotetext{\rosFormatOptions{2}{b}}{Spin period.}
\tablenotetext{\rosFormatOptions{3}{c}}{Spin-down rate.}
\tablenotetext{\rosFormatOptions{4}{d}}{Orbital period.}
\tablenotetext{\rosFormatOptions{5}{e}}{Estimated mass of visible pulsar.}
\tablenotetext{\rosFormatOptions{6}f}{Estimated mass of the pulsar companion.} 
\tablenotetext{\rosFormatOptions{7}g}{Eccentricity.}
\tablenotetext{\rosFormatOptions{8}h}{Beaming correction factor. An asterisk indicates the
beaming factor has not been experimentally determined; for these
systems, we use the canonical value $f_b=6$ (shown). }
\tablenotetext{\rosFormatOptions{9}i}{Characteristic age of the pulsar.}
\tablenotetext{\rosFormatOptions{10}j}{Spin-down age of the pulsar. We calculate $\tau_{\rm sd}$ only for recycled pulsars.}
\tablenotetext{\rosFormatOptions{11}k}{Merging time of a binary system due to the emission of gravitational waves.}
\tablenotetext{\rosFormatOptions{12}l}{Death time of the pulsar.}
\tablenotetext{\rosFormatOptions{13}m}{Most probable value of the total number of pulsars in a model galaxy estimated for the reference model (model 6 in KKL).}
\tablenotetext{\rosFormatOptions{14}n}{Parameter in rate equation [see Eq.~(\ref{eq:A})].}
\tablenotetext{\rosFormatOptions{15}o}{References: 
 (1) \cite{HulseTaylor:1975}; 
 (2) \cite{2000ApJ...528..401W};
 (3) \cite{1991Natur.350..688W}; 
 (4) \cite{2002ApJ...581..501S};
 (5) \cite{2004PhRvL..93n1101S};
 (6) \cite{psr:discovery:J0737};
 (7) \cite{1991Natur.350..688W};
 (8) \cite{psr:discovery:J1756-2251};
 (9) \cite{psr:discovery:J1811-1736};
 (10) \cite{2003MNRAS.342.1299K};  
 (11) \cite{psr:discovery:J1518+4904};
 (12) \cite{psr:discovery:J1829+2456};
 (13) \cite{psr:discovery:J0751+1807};
 (14) \cite{2005ApJ...634.1242N};
 (15) \cite{psr:discovery:J1157-5112}; 
 (16) \cite{psr:discovery:J1141-6545};
 (17) \cite{2003ApJ...595L..49B};
 (18) \cite{psr:discovery:B2303+46};
 (19) \cite{1999ApJ...516L..25V};
\nocite{psr:measurement:J1713+0747}
\nocite{psr:measurement:J0621+1002}
}
\rosFormatApJ{\end{deluxetable}}
\rosFormatGeneral{\end{deluxetable*}}

\subsection{Methodology I: Preferred population model}
In the context of \nsns,  KKL developed a statistical method to
estimate the likelihood 
of DCO formation rates, given an observed sample of DCOs in which one
member is a pulsar, designed to account
for the small number of known systems and their associated
uncertainties.  
Their analysis assumes that each binary is a unique representative of
its own class.   For each primitive class $\nu$ (i.e., for each pulsar), they find 
a posterior probability distribution
function ${\cal P}_\nu$ for its formation rate ${\cal R}_\nu$:
 \begin{equation}
 \label{eq:coreDistribution}
  {\cal P}_\nu({\cal R}) = A_\nu^2 {\cal R} e^{-A_\nu {\cal R}}.
 \end{equation}
The parameter $A_\nu$ depends on some of the properties of the pulsars
in the observed DCO  sample [see KKL Eq.~(17)]:
\begin{equation}
\label{eq:A}
A  = \tlife /(f_b \npsr)
\end{equation}
where $f_b^{-1}$ is the fraction of all solid angle the pulsar beam
subtends; $\tlife$ is the total \emph{detectable} lifetime of the object; and $\npsr$ is the total estimated number of systems
similar to each of the observed ones (i.e., $\npsr^{-1}$ is
effectively a volume-weighted probability that a pulsar with the same
orbit and an optimally oriented beam would be seen with any of the radio pulsar 
surveys; this factor incorporates all our knowledge of pulsar survey
selection effects as well as the pulsar space and luminosity
distributions).
In this section, we present  results only for our
preferred pulsar luminosity distribution model,  corresponding a 
power law luminosity distribution 
with negative slope, index $p=2$, and minimum ``luminosity\footnote{%
More correctly, this quantitiy represents pseudo-luminosity using a
$400\unit{Mhz}$ observing frequency.
}'' $L_{min}=0.3$~mJy kpc$^2$ 
[model 6 of  KKL; as discussed therein, this model
better accounts for more recent observations of faint pulsars].  In the
following section, we describe how our predictions change when
systematic uncertainties in $p$ and $L_{min}$ are incorporated into
${\cal P}$.

We obtain $\npsr$ via detailed simulations of pulsar survey selection
effects, as described in KKL
\citep[see also KKL and][]{Chunglee-wdns-proceedings}.
We obtain the beaming correction factor $f_b$
via direct pulsar measurements \citep{KNST} for those systems for
which data is available.  In the cases
where $f_b$ is not known, we choose a canonical value $f_b=6$ consistent with
existing measurements.
Finally, we estimate the detectable lifetime $\tlife$ via pulsar and
binary properties.
The relevant lifetime for  binaries detected via a
 pulsar component is
\begin{equation}
\label{eq:tau}
\tlife = \min(\tau_{sd},10{\rm Gyr}) + \min(\tau_d,\tau_{mrg})
\end{equation}
(where $\tau_{sd}$ is the pulsar spindown age [see \citet{Ar}]; $\tau_{mrg}$ is the time
remaining until the binary merges through the emission of
gravitational waves [see \citet{Peters:1964} and \citet{PetersMathews:1963}];
and $\tau_d$ is the pulsar ``death time'' [Eq. (9) from \citet{ChenRuderman:1993}]).
This calculation assumes the binary will be detectable via pulsar
emission from its birth (estimated by the spindown contribution) up
until its eventual merger via the emission of gravitational waves
(estimated via $\tau_{mrg}$) or until the pulsar ceases to emit
(estimated by $\tau_{d}$).
[Since $\tau_{sd}$ estimates are somewhat uncertain, we require that
they do not exceed the current age of the Galactic disk ($10$\,Gyr).] 
Table~\ref{tab:dcos}  presents appropriate values for $f_b$,
$\npsr$, $\tau_{\text{life}}$, and $A$ for each individual
pulsar-containing DCO.

Using these individual rate estimates, we then generate a distribution
for each \emph{net} formation rate for any class $K$ by assuming our individual
representatives are \emph{complete} as well as independent -- in other
words, we assume any any member of class $K$ (for $K=$NS-NS,  any
merging double NS) is similar to
one of the observed members of that class.  [Lacking
definitive information about orbital and pulsar parameter
distributions, we prefer not to perform unwarranted extrapolations 
to account for binaries that have not been seen.]  For example, to
estimate the formation rate of merging \nsns binaries in Milky
Way-equivalent galaxies (MWEG) , we take
 three merging \nsns binaries and combine their formation rate via 
\begin{eqnarray}
\label{eq:dcos:rates:convolve}
{\cal P}_{\text{\nsns}}({\cal R}_{tot}) &=& \int d{\cal R}_1 d{\cal R}_3 d{\cal R}_2
\delta ({\cal R}_{tot} - {\cal R}_1 -{\cal R}_2 -{\cal R}_3) \nonumber
\\
 &\times&  {\cal P}_1({\cal R}_1) {\cal P}_2({\cal R}_2) {\cal P}_3({\cal R}_3)
\end{eqnarray}
[This expression is discussed in \S 5.2 of KKL 
and presented explicitly for the three-binary case in Eq.~(A8) of
\citet{Chunglee-wdns-1}.]

Finally, for each class $K$ and thus each class rate distribution
${\cal P}_{K}$ we define symmetric 95\% confidence intervals:
 the upper and lower rate limits ${\cal R}_{w,\pm}$ satisfy
\begin{equation}
\int_0^{{\cal R}_{w,-}} d {\cal R} {\cal P}_w({\cal R}) =
\int_{{\cal R}_{w,+}}^\infty d {\cal R} {\cal P}_w({\cal R}) = 0.025
\; .
\end{equation}
This confidence-interval convention is different from  the customary
choice presented in \abbrvChungleeNSNSa.

In all plots that follow, rate probability distributions are
represented using a logarithmic scale for ${\cal R}$; thus, instead of
plotting ${\cal P}$, all plots instead show
\[
p(\log {\cal R}) = {\cal P}({\cal R}) {\cal R} \ln 10 \; .
\]

\begin{deluxetable}{lrrrr}
\tablecolumns{5}
\tablecaption{Constraint intervals}
\tablehead{\colhead{Type} & \colhead{Low} &
  \colhead{High}&
  \colhead{$p_{\rm ok}^a$}& \colhead{$p_{\rm ok+}^b$} \\
\colhead{} & \colhead{$\log_{10}(R {\rm yr})$} &
\colhead{$\log_{10}(R {\rm yr})$} &
}
\startdata%
NS-NS(m)  & $-4.50$ & $-3.65$ & 0.26 & 0.40\\
NS-NS(vw)& $-6.62$ & $-5.71$ &  0.36 & 0.60\\
WD-NS(e)  & $-4.94$ & $-3.98$ & 0.45 & 0.65\\
WD-NS(m)  & $-4.70$ & $-3.91$ & 0.35 & 0.60\\
SN Ib/c   & $-3.34$ & $-1.97$ & 1   & 1\\
SN II      & $-2.60$ & $-1.55$ & 1   & 1
\enddata
\label{tab:constraintIntervals}
\tablenotetext{\rosFormatOptions{1}a}{Fraction of population synthesis model parameters in
  our canonical model space 
  which produce rates in the constraint-satisfying interval.  See \S ~\ref{sec:ps}.}
\tablenotetext{\rosFormatOptions{1}b}{Fraction of population synthesis model parameters
  which produce rates within  the constraint-satisfying
  interval, modulo an estimate of the rms systematic error  associated with our
  fitting process. See \S~\ref{sec:ps} and 
\abbrvPSutiltwo.
}
\end{deluxetable}

\subsection{WD-NS binaries}
Four WD-NS binaries with relatively massive WDs have been discovered in the galactic disk: PSRs
J0751+1807, J1757-5322, J1141-6545, and B2303+46.  The intrinsic and deduced properties of
these four \wdns binaries are summarized in Table~\ref{tab:dcos}.  While all four
binaries may be applied to a \emph{net} WD-NS formation rate estimate,
the sample manifestly contains relics of distinctly different
evolutionary channels; for example, while J0751+1807 and J1757-5332
have evidently been strongly circularized and spun up by a recent mass transfer
episode, J1141-6545 cannot have been  \citep{psr:discovery:J1141-6545,2003ApJ...595L..49B}.  
Ideally, population synthesis must produce distributions of WD-NS
binaries consistent with both the observed orbital parameters and
spins.  The present sparse sample, however, does not allow a reliable
nonparametric estimate of the distribution of WD-NS binary
parameters.  Instead, as first step towards applying constraints based
on binary parameter \emph{distributions}, we
subdivide these four binaries into two overlapping classes:
\emph{merging} binaries, denoted WD-NS(m), which will merge through the
emission of gravitational waves within $10$ Gyr; and  \emph{eccentric}
binaries, denoted WD-NS(e), which have significant ($e>0.1$)
eccentricity at present.    The rate estimate derived for  both classes is
dominated by J1141-6545  \citep{Chunglee-wdns-1,Chunglee-wdns-proceedings}.  
Ignoring  systematic
uncertainties (e.g., assuming the beaming correction factor $f_b$ is
indeed its canonical value and not accounting the pulsar luminosity function uncertainties), Table~\ref{tab:constraintIntervals} shows the 95\% confidence
interval for each class' formation rate, shown per year for a
  Milky Way-like galaxy.

\subsection{\nsns\ binaries}
\label{sec:dco:nsns}
Seven \nsns binaries have been discovered so far in the Galactic disk.
Four of the known systems will have merged within $10$\,Gyr (i.e.,
``merging'' binaries: PSRs J0737-3039A, B1913+16, B1534+12, and
J1756-2251) and three are wide with much longer merger times (PSRs
J1811-1736, J1518+4904, and J1829+2456). The intrinsic and deduced properties of these seven
\nsns binaries are summarized in Table~\ref{tab:dcos}.  PSR~J1756-2251 was discovered
recently with acceleration searches \citep{psr:discovery:J1756-2251}.  
Since its pulsar and orbital parameters are  sufficiently similar to
PSR~B1913+16 and since the selection bias due to acceleration was
already included in KKL, we do not expect this new system to
significantly affect the NS-NS inspiral rate estimates.  As discussed
in more detail by \cite{2006.astro-ph..0608280}, the rate increase is
estimated to be smaller than 4\%.   For this reason, we omit it when
estimating NS-NS merger rates.

The observed NS-NS population naturally subdivides into two distinct classes, 
 depending on whether
they merge due to the emission of gravitational waves within 10 Gyr:
\emph{merging} \nsns binaries, denoted NSNS(m), and \emph{wide}
\nsns binaries, denoted NSNS(vw).   Table~\ref{tab:constraintIntervals} shows the 95\% confidence   
interval for each class' formation rate.

\subsection{Recycling, selection effects,  and the lack of wide NS-NS
  binaries}
\label{sec:dcos:recycling}
As Table~\ref{tab:constraintIntervals} indicates, the confidence
intervals for  wide and merging NS-NS binaries are almost an order of
magnitude from overlapping.  On the contrary, 
population
synthesis simulations produce merging and wide binaries at a roughly
equal (and always highly correlated) rates.  Since our rate estimation
technique automatically compensates for the most obvious selection
effects (e.g., orientation and detectable lifetime),  two
unbiased samples of wide and merging NS-NS binaries should arrive at
nearly the \emph{same} prediction for the NS-NS formation rate.

\abbrvPSconstraints{} 
explained this significant discrepancy by a selection effect:
evolutionary tracks leading to wide NS-NS binaries should be less likely to
recycle the first-born pulsar.  The \emph{observed} sample of wide
NS-NS binaries [NSNS(vw)] represents a much smaller subset of
\emph{recycled} pulsars.   As summarized in 
\abbrvPSutiltwo,
we assume any wide NS-NS binary which in its past underwent
\emph{any conventional}, i.e., dynamically stable, mass transfer  will recycle one of its neutron stars to a long-lived pulsar. 
This  condition is likely to over-estimate  the true
NSNS(vw) formation rate. We note that neutron-star recycling can also possibly happen during dynamically unstable mass-transfer phases (common envelopes), but the vast majority of wide NS-NS do not evolve through such a phase. 

\subsection{Pulsar Population Model Uncertainties}
As noted in KKL and subsequent papers, our reconstruction of the
pulsar population (i.e., $N_{PSR}$) relies upon our understanding of
pulsar survey selection effects and thus on the underlying pulsar luminosity
distribution.  This distribution can be well-constrained
experimentally \citep[see, e.g.][]{1997ApJ...482..971C}, though these
constraints do not yet incorporate  recent
faint pulsar discoveries such as PSR J1124-5916
\citep{2002ApJ...567L..71C}. 
Nonetheless, different observationally-consistent
distributions imply significantly different distributions, with
maximum-likelihood rates differing by factors of order $10$
(\abbrvChungleeNSNSa). 
Since the constraint intervals discussed above assume the \emph{preferred pulsar
  luminosity distribution model} --  a power-law pulsar luminosity
distribution with negative slope $p=2$ and minimum cutoff luminosity
$L_{min}=0.3$ mJy kpc$^2$
-- they do not incorporate any uncertainty in the pulsar luminosity function.

The infrastructure needed to incorporate uncertainties in the pulsar
luminosity function has been presented and applied, for example in
\citet{2006.astro-ph..0608280}. 
However, out-of-date constraints on the pulsar luminosity function
allow implausibly high minimum
pulsar luminosities $L_{min}$.  A high minimum pulsar luminosity
implies fewer merging pulsars have been missed by surveys.  Thus, these
out-of-date constraints on pulsar luminosity functions permit models
consistent with 
substantially lower merger rates than now seem likely, given the
discovery of faint pulsars.  In other words, if we use the infrastructure
presented in \citet{2006.astro-ph..0608280} to  marginalize over $L_{min}$
and $p$ generate a \emph{net} distribution function for the merger
rate, then the 95\% confidence intervals associated with that net
distribution would have a spuriously small lower bound, entirely
because the pulsar population model permits large $L_{min}$.
Therefore, in the present study we present results based only on our
preferred luminosity function and do not include the out-of-date
luminosity function constraints by \cite{1997ApJ...482..971C} and the
related net rate distribution provided by \citet{2006.astro-ph..0608280}.


\section{Observations of supernovae}
\label{sec:data:other}
Type Ib/c and II supernovae occur extremely rarely near the Milky Way.  
While historical data contains several observations of and even
surveys for supernovae, the selection effects in these long-duration
heterogeneous data sets have made their interpretation difficult
(Cappellaro 2005, private communication).  In this paper, we estimate
supernova rates and uncertainties using Table 4 of
\citet{Cappellaro:SNa}. 
  To translate this table, which expresses results and their
  (one-sigma, systematics-dominated) uncertainties
in terms of  a number of supernovae per century per $10^{10}$ blue solar
luminosities ($L_{\odot,B}$), into the equivalent rates  per Milky Way
equivalent galaxy, we 
assume a Milky Way blue luminosity of 
$1.7\times 10^{10}L_{\odot,B}=0.9\times 10^{10} L_{\odot}$ relative to
the solar blue $(L_{\odot,B})$ and total $(L_\odot)$ luminosities
(see, e.g., 
\citet{KNST},
\citet{1991ApJ...380L..17P}, \citet{ADM:Cox2000},
and references therein).   More specifically, we translate the
  supernovae rates that appear in their Table 4, expressed in ``SNU'' (1 SNU =
$1/10^{10}L_{\odot,B}\times 10^2{\rm yr}$) into a
supernovae rate expressed in events per year per Milky Way-equivalent
galaxy (MWEG) using a proportionality constant 
$
W = 2\times 10^{-2} {\rm MWEG}/  {\rm SNU} 
$.
For simplicity, we ignore $O(30\%)$ uncertainty in the blue luminosity of the
Milky Way and thus $W$ \citep[see][for an estimate of this uncertainty]{KNST}.  

Unfortunately, without  substantially more detailed information on and
statistics of the limiting (systematic) errors, we cannot
justify any relation between the \emph{linear, one-sigma} confidence intervals provided in
their Table 4 and the 
\emph{logarithmic, two-sigma} 
confidence intervals needed in our analysis.   Lacking any way to
rigorously justify a \emph{useful} choice\footnote{In fact, the Chebyshev inequality $P(|x-\mu|>t)\le
  \sigma^2/t^2$, which applies to any random variable with mean $\mu$ and standard deviation
  $\sigma$, provides a universal tool with which we can find a
  better-than-$95\%$ ``overconfidence interval'' on  $r$ and thus $\log
  R$.   However, not only does that interval in $r$
  includes $r=0$ -- so 
  arbitrarily small supernovae rates would be included -- but it also
  corresponds to an excessively conservative $4.5\sigma$ confidence
  interval for Gaussian-distributed errors.} 
for a bounded two-sided confidence
interval in $\log R$,
we decided to employ the logarithmic confidence interval 
\begin{eqnarray}
\log (r W) \pm 2 \, {\rm max}\left( 
  \left|\log \frac{r+u}{r}\right|,
  \left|  \log \frac{r-u}{r}\right|
  \right)
\end{eqnarray}
where $r$ and $u$ are the average value of and one-sigma uncertainty in the
supernovae rate (in SNU); the corresponding explicit confidence
intervals (in $\log R$ per year per MWEG) for
Type II and Ib/c supernovae are shown both in Table
\ref{tab:constraintIntervals} and Figure \ref{fig:ConstraintsVersusDistribs:SN}.
Intuitively, this logarithmic interval arises from (i) converting the mean
value and limiting bounds of a one-sigma confidence interval in $r$ to
a ``one-sigma'' confidence interval in $\log R$; (ii) calculating the
maximum distance from the central value to the corresponding two
``one-sigma'' limits in $\log R$;  and then (iii) doubling that
distance to generate a ``two-sigma'' interval.
[A marginally more optimistic case using the \emph{minimum} distance
from center to each limit also yields nearly the same intervals.]



Though fairly accurate studies exist of the high-redshift
supernova rate \citep[e.g.,][]{Cappellaro:SN-highz}, they have little  relevance
to the present-day Milky Way.
Several surveys
have also attempted to determine the supernova rate in the Milky Way by a variety of
indirect methods, such as statistics of supernova remnants
\citep[highly unreliable due to  challenging selection effects; see, e.g.,][]{1991ARAA..29..363V}
and direct observation of radioisotope-produced backgrounds \citep[e.g., decay from ${}^{26}Al$, as described
in][]{2006Natur.439...45D}.  Taken independently, these methods
have greater uncertainties than the historical studies of
\citet{Cappellaro:SNa}.

\section{Population synthesis predictions}
\label{sec:ps}

\subsection{StarTrack population synthesis code}
We estimate formation and merger rates for several classes of double
compact objects using the 
\emph{StarTrack} code first developed by BKB 
 and recently significantly updated and tested as described in detail in
\citet{StarTrack2};
see \S 2 of \citet{ChrisBH2007}
for a succinct description of the changes between versions.   
This updated code predicts somewhat different double compact object
properties than the version used in BKB.

Additionally, 
 we have adopted a maximum
neutron star mass of $2.5 M_\odot$ (raised from $2 M_\odot$ used in
our two previous studies \abbrvPSconstraints{}  and \cite{PSutility}, henceforth \abbrvPSutility),  motivated by recent observations suggesting  pulsars
with masses near $2 M_\odot$ (see, e.g.,\citet{2005ApJ...634.1242N} and
\citet{psr:measurement:J0621+1002}).
Such a higher maximum 
neutron star mass  effectively converts many merging binaries we would otherwise
interpret as BH-NS or even  BH-BH binaries into merging NS
binaries.   Consequently the BH-NS and particularly BH-BH rates are
decreased
from the values seen in \abbrvPSconstraints.  [As a concrete
example, with a low maximum NS mass, many BH-NS binaries form through
accretion-induced collapse of a NS to a BH, as seen in Figure 1 of
\cite{2005ApJ...632.1035O}.  However, a significant fraction of these
systems would be interpreted as NS-NS binaries with the adopted higher maximum
NS mass; see \cite{ChrisBH2007} for a detailed discussion to merger
rates that 
this revised minimum neutron star mass implies.%
] 

Like any population synthesis code, it evolves randomly chosen
binaries from their birth to the present, tracking the stellar and
binary parameters.  For any  class of events that is
\emph{identifiable within the code}, such as supernovae or DCO
mergers, we estimate event rates by taking the average event rate
within the simulation (i.e., by dividing the
total number of events seen within some simulation by the duration of
that simulation) and renormalizing by a scale factor that depends on
properties of the simulation (i.e., the number of binaries simulated
and the binary birth mass distributions assumed) and the Milky Way as
a whole (i.e., the present-day star formation rate). 
  Specifically, our simulations are normalized to be consistent with
  an assumed Milky Way star formation rate\footnote{Our approach gives only the \emph{average} event rate.  The present-day merger rate agrees with this quantity when most mergers occur relatively promptly (e.g., $<100$ Myr) after their birth. Some DCOs -- notably double BH binaries --  have substantial delays between birth and merger, introducing a strong time dependence to the merger rate.  The technique described above will significantly
\emph{underestimate} these rates.  This point will be addressed in
  considerably more detail, both for the Milky Way and for a
  heterogeneous galaxy population in forthcoming papers
  by
  \citet{PSgrbs-popsyn,PSellipticals}. 
} of $3.5 M_\odot {\rm yr}^{-1}$.  The necessary formulae are presented and discussed extensively in
a companion paper \abbrvPSutiltwo.

Finally, like any population synthesis code, the relative
probabilities of different evolutionary endpoints are heavily
influenced by the relative likelihood of different initial
conditions, such as the distribution of initial semimajor axes.  
For example, observational data is consistent with initial semimajor
axes distributed uniformly in $\log a$; see e.g., Figure 2 of
\cite{1983ARAA..21..343A} and Figure 7 of
\cite{1991AA...248..485D}.\footnote{The semimajor axis distributions
  in these two papers are relatively similar, though they do predict a
  slightly different proportion of binaries in any small bin
  $dn/da$.  However, for spiral galaxies the instantaneous merger rate 
  depends on the fraction of binaries inside a \emph{broad} period
  interval, not the derivative at any point.}
 Considering the limits of the period distribution and the masses
involved, we adopt a semimajor axis distribution flat in $\log a$ from
contact to $10^5 R_\odot$ (see \abbrvStarTrack). 

When constructing our archived population synthesis results, we did
not choose to record detailed information about the nature and amount
of any mass transfer onto the first-born NS.  We therefore cannot
reconstruct precise population synthesis predictions for the NSNS(vw)
formation rate.  However, we do record whether some mass transfer
occurs, and the nature of the mass transfer mechanism.  The
conventional mass transfer mechanism -- dynamically stable Roche-lobe
overflow -- inevitably produces a disk around the compact  object and
can potentially spin that object up.   Other mechanisms, such as
(possibly hypercritical) common-envelope evolution, presumably involve
a substantially more spherical accretion flow; the specific angular
momentum accreted may be substantially lower, possibly not enough to
recycle the neutron star.  Thus for the purposes of
identifying a class of potentially recycled (``visible'') wide NS-NS
binaries, we assume any system which underwent mass transfer  (dynamically stable in the case of wide systems) produced a recycled NS primary. 


\subsection{Fitting results}
As in previous studies such as \abbrvPSutility, 
here we have chosen to vary seven (7)
model parameters in the synthesis calculations.  These choices are strongly guided by our past
experience with double-compact-object population synthesis 
and represent the model parameters for which strong dependence has been confirmed. 
These seven parameters enter into every aspect of our population
synthesis model.  One parameter, a power law index $r$ in our
parameterization of the companion mass distribution, influences 
the initial binary parameter distributions (through the companion
masses).   Another parameter $w$, the massive stellar wind strength,
controls how rapidly the massive progenitors of compact objects lose
mass; this parameter strongly affects the final compact-object mass
distribution.  Three parameters $v_{1,2}$ and $s$ are used to
parameterize the supernovae kick distribution as a superposition of
two independent maxwellians.   These kicks provide critical opportunities to
push distant stars into tight orbits and also to disrupt
  potential double neutron star progenitors.    Finally, two parameters
$\alpha \lambda$ and $f_a$ govern  energy and mass
transfer during certain types of binary  interactions; see Section 2.2.4 of
\abbrvStarTrack{} 
for details.

Other parameters,
such as the fraction of stellar systems which are binary (here,
we assume a binary fraction of 100\%) and the distribution of initial binary parameters,
are comparatively well-determined  (see e.g..\cite{1983ARAA..21..343A},
\cite{1991AA...248..485D} and references
therein); for these parameters we have adopted preferred values.\footnote{Particularly for the application at hand -- the
  gravitational-wave-dominated delay between binary birth and merger
  -- the details of the semimajor axis distribution matter little.
  For a similar but more extreme case, see \cite{clusters-2005}.}
%
Even for the less-well-constrained parameters,  some inferences 
have been drawn from previous
studies, either more or less directly (e.g., via observations of
pulsar proper motions, which  presumably relate closely to supernovae
kick strengths; see, e.g., \citet{HobbsKicks}, \citet{ArKicks},
\citet{2006ApJ...643..332F} and references therein)
or via comparison of  some subset of  binary population synthesis
results with observations 
(e.g.,  \S 8 of \citet{StarTrack2}, \citet{2006AA...460..209V},
\citet{2005MNRAS.356..753N}, \citet{2002MNRAS.337.1004W},
\citet{2002ApJ...565.1107P} and references therein).
Despite observational suggestions that point towards
preferred values 
for these seven parameters -- and despite the good agreement with short
GRB and other observations obtained when using these preferred values
(\citet{ChrisShortGRBs}) -- in this paper we will \emph{conservatively} examine the
implications of a \emph{plausible range} of each of these parameters.
More specifically, despite the Milky Way-specific studies of
\citet{PSconstraints,PSmoreconstraints} (which apply only to spirals,
not the elliptical galaxies included in this paper), 
to allow for a broad range of possible models, we consider the specific
parameter ranges quoted in \S 2 of \citet{PSutility} for all
parameters except kicks.   For supernova kick parameters, we allow the
dispersion $v_{1,2}$ of either component of a bimodal Maxwellian to run from 0
to 1000 km/s.

Since population synthesis calculations involve considerable
computational expense, in practice we estimate the merger rate we
expect for a given combination of population synthesis model
parameters via seven-dimensional fits to an archive of roughly 3000
detailed simulations, as presented and analyzed in
detail in \abbrvPSutiltwo{}.
However, these fits
introduce systematic errors, which have the potential to
significantly change the predicted set of constraint-satisfying
models.  For this reason, \abbrvPSutiltwo{} also 
estimates the rms error associated with each seven-dimensional fit.
Finally, \abbrvPSutiltwo{} demonstrates that, by broadening the constraint-satisfying
interval by the fit's rms error, 
the predicted set of constraint-satisfying models includes most models
which \emph{actually} satisfy the constraints.  For this reason, the
dark shaded regions in Figures
\ref{fig:ConstraintsVersusDistribs:SN},
\ref{fig:ConstraintsVersusDistribs:WDNS}, and
\ref{fig:ConstraintsVersusDistribs:NSNS}] account for both
observational uncertainty in the Milky Way merger rate and for
systematic errors in the fits against which these observations will be compared.

\subsection{Prior distributions}

\begin{figure}
 \includegraphics[width=\columnwidth]{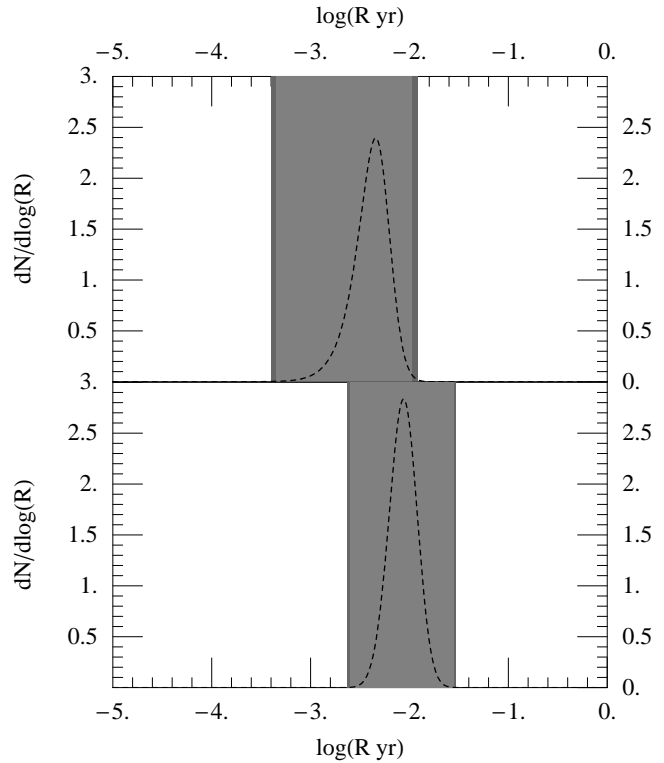}
\caption{\label{fig:ConstraintsVersusDistribs:SN} Probability
  distributions for SN Ib/c (top) and SN II (bottom), as predicted by
  population syntheses with \emph{StarTrack}.  The light shaded region
  shows the ($\approx 2\sigma$) interval
  consistent with observational constraints of supernovae, from
  \citet{Cappellaro:SNa}.  A dark shaded region (barely visible in this
  plot) indicates these constraints, augmented by an estimate of the
  systematic errors in our fits; see \abbrvPSutiltwo.
These constraints provide no information about
  population synthesis.
}
\end{figure}

\begin{figure}
\includegraphics[width=\columnwidth]{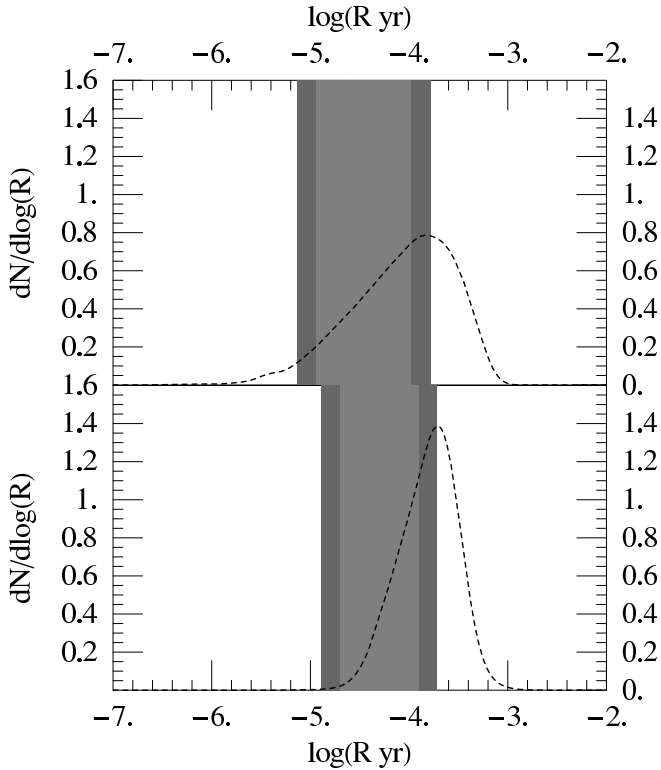}
\caption{\label{fig:ConstraintsVersusDistribs:WDNS} Probability
  distributions for merging (top) and eccentric (bottom) WD-NS
  formation rates, as predicted by
  population synthesis.  The light shaded region shows the 95\% confidence interval
  consistent with observations of binary pulsars, as described in
  Sec.~\ref{sec:data:DCOs}.  The dark shaded region extends this
  constraint interval by 
an estimate of the
  systematic error in each fit; see \abbrvPSutiltwo.
}
\end{figure}

\begin{figure}
\includegraphics[width=\columnwidth]{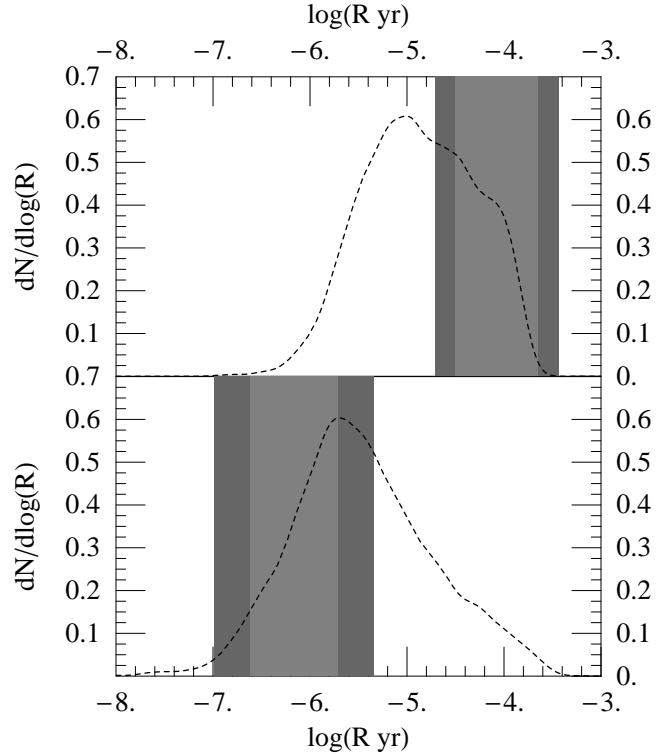}
\caption{\label{fig:ConstraintsVersusDistribs:NSNS} 
Probability
  distributions for merging (top) and wide (bottom) NS-NS
  formation rates, as predicted by
  population synthesis.  The light shaded region shows the 95\% confidence interval
  consistent with observations of binary pulsars, as described in
  Sec.~\ref{sec:data:DCOs}.
Because of sparse sampling and
  relatively poor data, our fit for the
  formation rate of visible, wide NS-NS binaries is significantly more
  uncertain than other fits.
}
\end{figure}

Lacking knowledge about which population synthesis model is correct,
we assume all population synthesis model parameters in our range are
\emph{equally likely}.  A  monte carlo over the seven-dimensional
parameter space allows us then to estimate the relative likelihood,
all things being equal, of various DCO merger rates (shown in 
Figures~\ref{fig:ConstraintsVersusDistribs:WDNS},\ref{fig:ConstraintsVersusDistribs:NSNS},
and \ref{fig:FinalRateDistribs}) and supernova rates (Figure~
~\ref{fig:ConstraintsVersusDistribs:SN}).
Also shown in these figures are the observational constraints
described in Sections \ref{sec:data:DCOs} and \ref{sec:data:other} (shown in shaded gray).
Table~\ref{tab:constraintIntervals} provides both the explicit confidence intervals of the constraints and the fraction of \emph{a priori} models which satisfy that constraint.
Finally, the dotted lines in Figure~\ref{fig:FinalRateDistribs} show the
a priori population synthesis predictions for BH-NS and NS-NS
merger rates in the Milky Way.

\section{Applying and employing constraints}
\label{sec:constraints:DCO}

Physically consistent models must reproduce all predictions.
Supernovae rates, being ill-determined due to the large observational systematic
errors mentioned in Sec.~\ref{sec:data:other}, are easily reproduced
by almost all models at the $2\sigma$ level;
see Fig.~\ref{fig:ConstraintsVersusDistribs:SN}.  However, population
synthesis models do not always reproduce observed formation rates for
DCOs, as shown in Figs.~\ref{fig:ConstraintsVersusDistribs:WDNS} and
\ref{fig:ConstraintsVersusDistribs:NSNS}.  These four 95\% confidence
intervals implicitly define a $(0.95)^4\approx 81\%$
confidence interval in 
the seven-dimensional space, consisting of $9\%$ of the 
original parameter volume when systematic
errors in our fitting procedure are included; see
Table~\ref{tab:constraintIntervals}.\footnote{Our confidence in this
  interval is significantly higher than 81\%, because we have very
  conservatively incorporated estimates for systematic errors in our
  fits; see \abbrvPSutiltwo{} 
for details.}  In other words, we are
quite confident ($80\%$ chance) that the physically-appropriate
parameters entering into \emph{StarTrack} can be confined within a
small seven-dimensional volume, in principle significantly reducing
our model uncertainty.

\begin{figure}[h]
\includegraphics[width=\columnwidth]{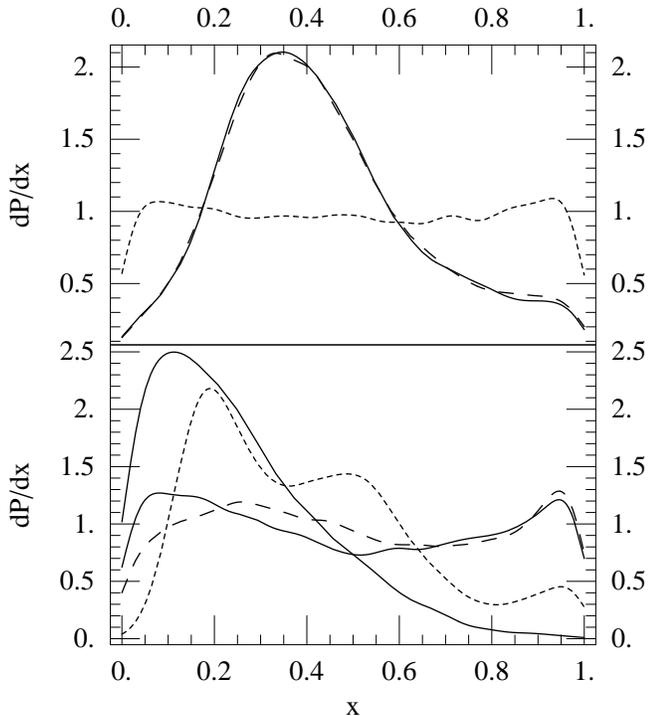}
\caption{\label{fig:params} Differential probability distributions
  $dP_k/dx$ defined so $P_k(x)$ is the fraction of all models
  consistent with all DCO observations and with 
  $x_k<X$.  The top panel shows the distributions for the 3
  kick-related parameters $x_3,x_4,x_5$ (dashed, solid, and dotted,
  respectively, corresponding to $v_1$, $v_2$, and $s$).  The bottom panel shows the
  distributions for  $x_1$ (the mass-ratio distribution parameter $r$, appearing as
  the solid nearly constant curve), and the three
  binary-evolution-related parameters $x_2,x_6,x_7$ (dashed, dotted,
  and solid, respectively, corresponding to $w$, $\alpha \lambda$, and
  $f_a$).
  The distribution of $f_a=x_7$ exhibits a strong peak somewhere
  between $0-0.1$.  Our choice of smoothing method causes all
  projected distributions to appear to drop to zero at the boundaries,
  as it involves averaging with empty cells.
}
\end{figure}

\begin{figure*}
\includegraphics[width=\textwidth]{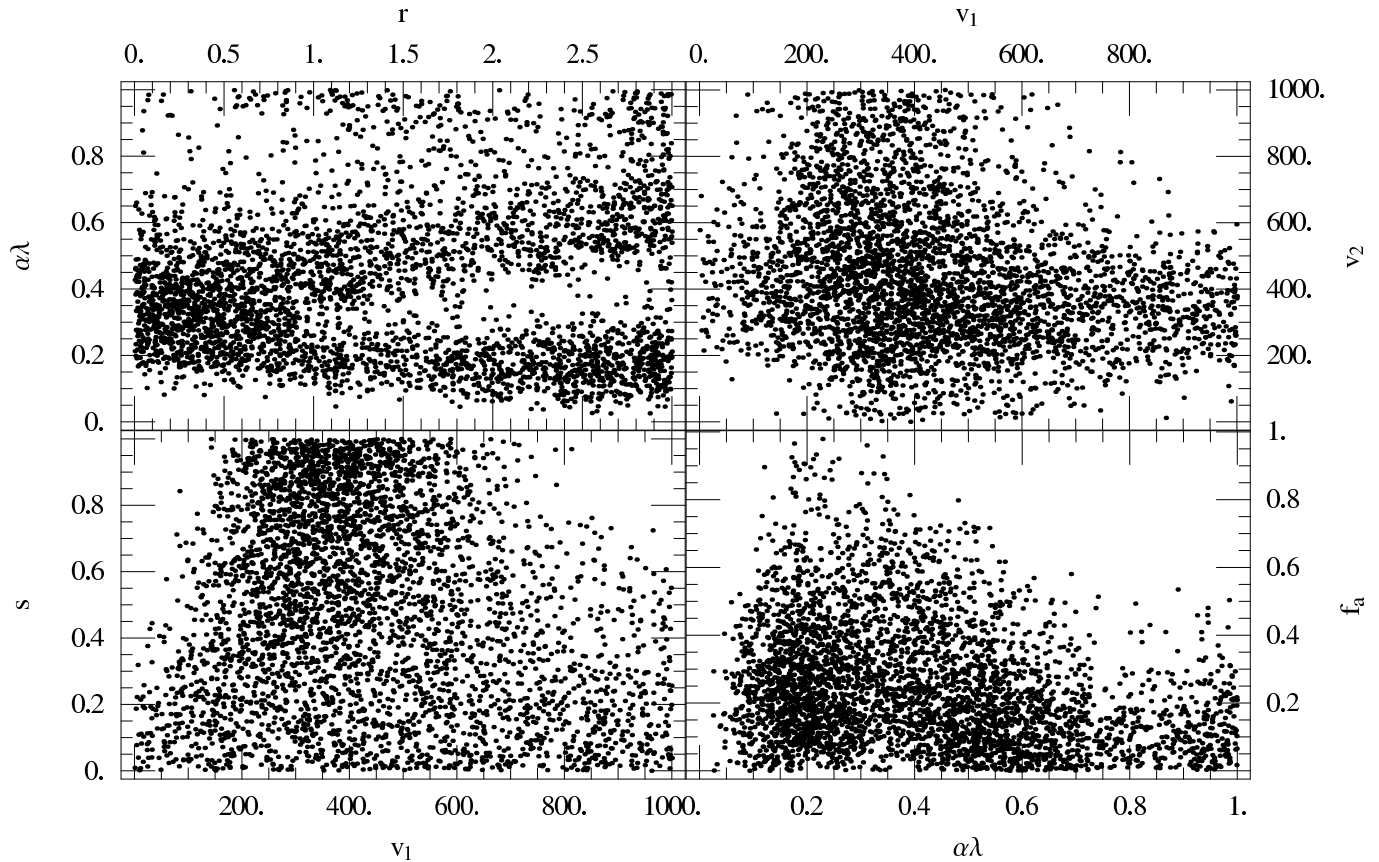}
\caption{\label{fig:params:correlations} For 4000 randomly chosen
  constraint-satisfying parameter combinations, scatter plots  of
  $(r,\alpha\lambda)$ (top left),
  $(v_1,v_2)$ (top right),  $(v_1,s)$ (bottom left) and $(\alpha \lambda,f_a)$
  (bottom right). 
} 
\end{figure*}

In Figure~\ref{fig:params} we show  one-dimensional projections onto
each coordinate axis of the constraint-satisfying volume -- in other
words, the distribution of values each \emph{individual} population
synthesis parameter can take.  For the
purposes of this and subsequent plots, we use the following seven
dimensionless parameters $x_k$ that run from 0 to 1: $x_1= r/3$; 
$x_2=w$; $x_3=v_1/(1000{\rm km/s})$, $x_4=v_2/(1000{\rm
  km/s})$; $x_5=s$; $x_6=\alpha\lambda$, and $x_7=f_a$.   
%
%
As seen in the top panel of this figure, the \emph{single most likely} distribution of supernovae kicks bears a surprising
resemblance to the observed pulsar kick distribution
\citep[see,e.g.][and references
therein]{Ar,HobbsKicks,2006ApJ...643..332F}, even though \emph{no
  information about pulsar motions have been included} among our
constraints and priors.  However, as is clear from Figure
\ref{fig:params} and more clearly by the top right panel of Figure
\ref{fig:params:correlations}, a wide range of kick velocity
distributions remain consistent with observations, including those
with very low ($\simeq 100\unit{km/s}$) and very high ($\simeq 600
\unit{km/s}$) characteristic velocities.    The population synthesis mass transfer
parameter $f_a$ is particularly well-constrained, with a strong maximum near
$x=0$,
implying that mass-loss fractions of about 90\% or higher are favored
in non-conservative, but stable, mass transfer episodes. Also, common
envelope efficiencies $\alpha\lambda$ of about  0.15-0.5 appear to be
favored by the constraints. The rest of the one-dimensional parameter
distributions are nearly constant and thus uninformative.   
Though small, the
constraint-satisfying volume extends throughout the seven-dimensional
parameter space.  
Higher-dimensional correlations disguise most of the remaining
information, as indicated in Figure \ref{fig:params:correlations}.


\begin{figure}
\includegraphics[width=\columnwidth]{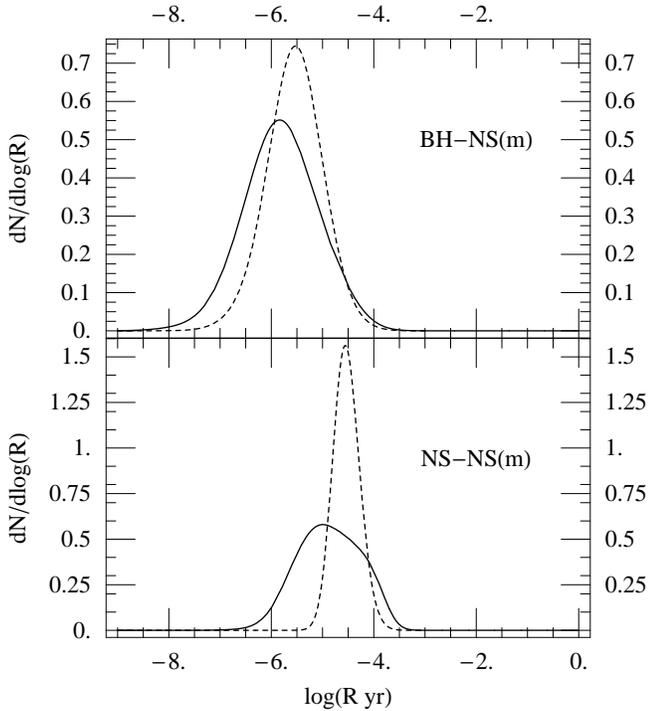}
\caption{\label{fig:FinalRateDistribs} A plot of the  probability distribution for the BH-NS (top) and NS-NS (bottom) merger rates per
Milky Way-equivalent galaxy, allowing for systematic errors
in the  BH-NS and NS-NS fits.   The solid
 curves result from smoothing a histogram of results from a random
 sample of population synthesis calculations with  systematic errors included.
The dashed  curve  shows
 the same results, assuming model parameters are restricted to those
 which satisfy all four DCO constraints (both WD-NS
and both NS-NS constraints) described in Sections \ref{sec:data:DCOs}
and \ref{sec:data:other} and summarized in
Table~\ref{tab:constraintIntervals}.   Compare with Figure 5 of
\citet{PSconstraints}, which uses substantially different priors.
}
\end{figure}

\begin{figure}
\includegraphics[width=\columnwidth]{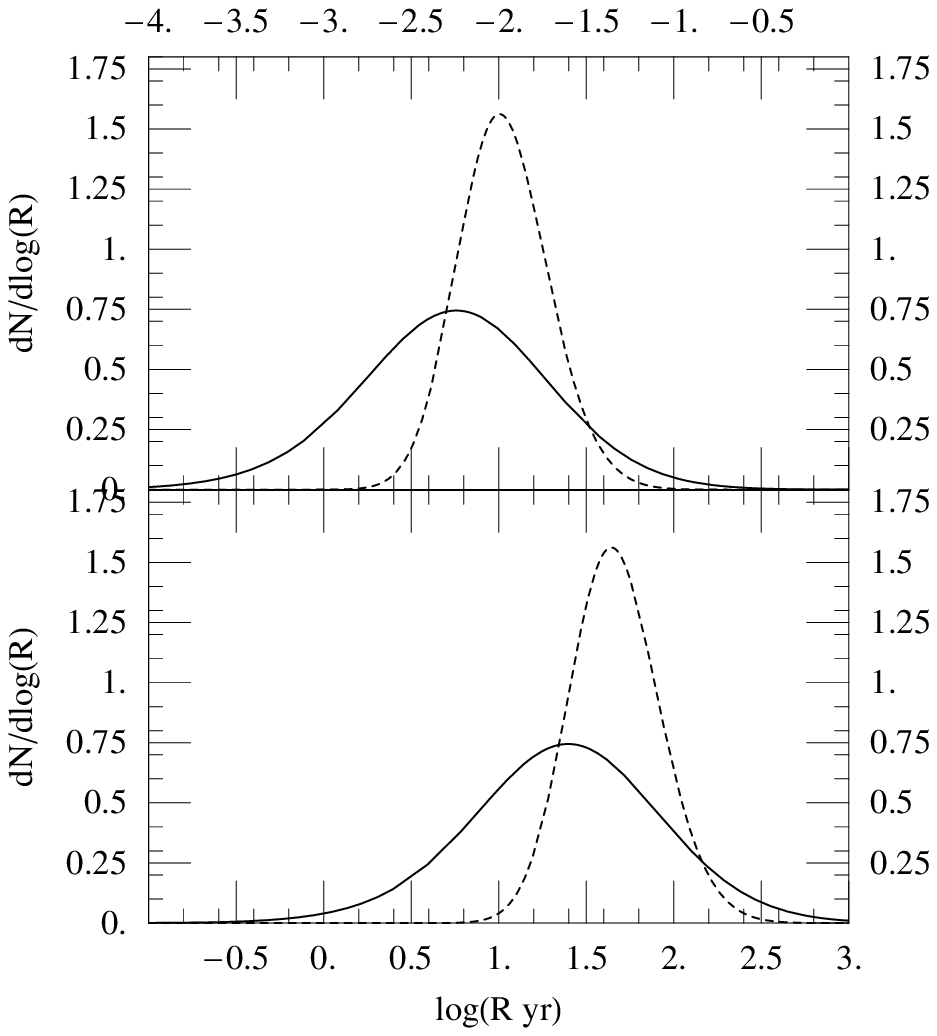}
\caption{\label{fig:results:LIGO1} Range of expected detection rates for
  initial (top) and advanced (bottom) LIGO for BH-NS (solid) and NS-NS (dotted)
  binaries, based on models that satisfy observational constraints and
  under the assumptions outlined in \citet{PSconstraints} (e.g., using
  a fixed chirp-mass spectrum and LIGO range for detection, as well as assuming Milky-Way like
  galaxies fill the universe with density $0.01 {\rm Mpc}^{-3}$). 
  These estimates incorporate systematic errors due to the fitting
  procedure.  
}
\end{figure}

These results should
be contrasted with the distinctly different post-constraint results
shown in \citet{PSconstraints} (cf. their Figure 5 and C7).  Notably, even
though \emph{two fewer} constraints were imposed, they find a
significantly smaller ($2\%$ versus $9\%$) constraint-satisfying
volume. 
 In large part, these
differences can be ascribed to including systematic
errors: in \citet{PSconstraints}, preliminary data for
the sparse and poor-quality NSNS(vw) sample was fit and applied without any
account for fit-induced errors.   However, our calculation also
differs at several fundamental levels from the original approach :
(i) the space of  models studied is larger, covering more area in
$v_1,v_2$ 
(see, e.g., \abbrvPSutiltwo);
(ii) the observed sample of merging NS-NS binaries is compared with
the \emph{total} merging NS-NS formation rate predicted from
population synthesis, rather than with subset of merging NS-NS which
undergo mass transfer; and critically (iii) the constraints are more numerous.

As in \citet{PSconstraints},  evaluating fits for other DCO merger
rates  on the constraint-satisfying
model parameters provides  revised estimates for various phenomena
of interest, such as for the  BH-NS and NS-NS merger rates.
Figure~\ref{fig:FinalRateDistribs} shows smoothed
histograms of the BH-NS and NS-NS  merger rates, both before and
after observational constraints were applied.    
To be conservative, these distributions also incorporate our best estimate
for systematic errors associated with fitting to the data: rather than simply showing the histogram of
merger rates that follows by evaluating the fits $\hat{R}(x)$ on many
randomly poppulation synthesis parameters $x$, we \emph{smooth} the
resulting histogram by a gaussian with width  $J_q$ (i.e., by the
characteristic systematic error).   
For BH-NS and NS-NS binaries this characteristic error is
significantly smaller than the characteristic width of the
distribution.  

Finally,  in Figure \ref{fig:results:LIGO1} we  estimate the range of initial
and advanced LIGO detection rates implied by these constrained
results.  This estimate combines our constrained population synthesis
results; our estimates for systematic fitting error; the approximate LIGO range as
a function of chirp mass $M_c$:
$
d = d_o \left({\cal M}_c/ 1.2 M_\odot \right)^{5/6}
$ where $d_o=15\sqrt{3/2}$, $300\, {\rm Mpc}$ for initial and advanced
LIGO, respectively;
 estimates for the average volume detected for a
given species, as represented by 
$\left<M_c^{15/6}\right>=11.2,2.3 M_\odot$ for BH-NS
and NS-NS, respectively  \citep[cf.][our average chirp masses are
systematically higher because of our higher maximum neutron star mass]{PSconstraints};
and a homogeneous,
Euclidean universe  populated with Milky Way-equivalent galaxies with
density $0.01 {\rm Mpc}^{-3}$, each forming stars at rate $3.5
M_\odot{\rm yr}^{-1}$  \citep{KNST}.  
%
The high rate of double neutron star mergers [NS-NS(m)] demanded by
observations explains most of the differences between  the
relative likelihoods of various merger rates based on \emph{ a
  priori} (solid) and constrained (dashed) expectations.  For example,
 the lowest
$\alpha\lambda<0.1$  eliminate too many
NS-NS binaries to be consistent with the many observed Galactic NS-NS binaries.
Additionally, the differences between this paper's results and those
found in \abbrvPSconstraints are in part produced from the different
\emph{assumptions} (priors) about the relative likelihood of different
population synthesis parameters: our distributions of constrained and
unconstrained merger rates differ from previous studies in large part
because this study allows a broader range of supernovae kicks.

We emphasize that the new formation and merger rates of double compact
objects that we use are  
the result of the revised population synthesis code \emph{StarTrack}
\citep{StarTrack2}. The underlying physics, and formation
scenarios of BH-NS and NS-NS binaries will be discussed in full detail in \citet{Chris-nspaper-2007}.  

\section{Conclusions}
\citet{PSconstraints}  performed a
proof-of-concept calculation to compare population synthesis
simulations with observations, interpret the resulting constrained
volume, and apply the constraint-satisfying parameters to revise a
priori predictions of astrophysically critical rate results.
In this paper, we demonstrate that even when 
systematic  model fitting errors are aggressively overcompensated  for, 
observational constraints still provide information about population
synthesis parameters.   These results include both surprisingly good
agreement with pulsar kick distributions and strong constraints on 
at least one
parameter involved in binary evolution.   Since in almost
all cases our systematic errors appear much smaller than those from
observations, we are confident that now \emph{observational} limitations,
rather than computational ones, primarily limit  our ability to constrain
population synthesis results.   In particular, we expect stringent
tests of  population synthesis models are possible, beginning by
imposing more  observational constraints than varied model parameters
(seven in our case).


Our analysis remains predicated upon a correct identification of all
parameters and input physics critical for the formation of double
compact objects. Admittedly, our understanding of binary evolution
continues to evolve; however multiple population studies from
different groups over the years lead to very similar conclusions about
the basic input physics that primarily affects the formation rates of
double compact
objects~\citep{StarTrack,2002MNRAS.329..897H,1996AA...309..179P,1998ApJ...496..333F}. In
this paper we use a relatively new version of the \emph{StarTrack}
code that is described in great detail in\cite{StarTrack2}. A
forthcoming paper \citep{Chris-nspaper-2007} will discuss how
these updates affect the evolutionary history of NS binaries 
in significantly greater detail. 
  
Our analysis also remains predicated upon a correct and complete
interpretation of 
the observational sample and associated systematic and model
uncertainties. In some respects, however, our models for observations 
and their biases may be incomplete or
not representative. 
We effectively assume pulsar recycling is required to detect binary
pulsars in our implicit hypothesis 
that only one pulsar in NS-NS systems, the one observed,
was ever likely to have been detected.    And finally we use a
canonical value for pulsar beaming correction factor $f_b$ for several
systems with known spins but unknown beaming geometry. 


And to be comprehensive regarding the factors neglected here, our
population synthesis parameter study has not varied over \emph{all}
plausible models.  We have neither considered purely polar supernova kicks
\citep[see][]{2005MNRAS.364.1397J} nor changed the maximum NS mass
from $2.5 M_\odot$ nor even allowed for a heterogeneous birth population of single stars 
and binaries (a binary fraction of 100\% is assumed in all models).  And of course we continue to use a single fixed star
formation rate for the Milky Way, rather than treat it as an
independent uncertain but constrained parameter.

However, all these limitations can be remedied by closer study.  
Thus, we  expect that either (i) all DCO merger
rates will become determined to within $O(50\%)$ or better (based on
the comparative accuracy to which we know NS-NS merger rates), or (ii)
experimental data will conflict with the prevailing notion of binary
population synthesis, revealing flaws and limitations in classical
parametric models for binary stellar evolution.

\acknowledgements
\begin{acknowledgements}
 We thank NCSA for providing us with resources used to perform many of the computations in this text.
This work is partially supported by NSF Gravitational Physics grant PHYS-0353111, and a David and Lucile Packard Foundation Fellowship in Science and Engineering to VK.  KB acknowledges support from the Polish Science Foundation (KBN) Grant 1P03D02228.
\end{acknowledgements}

\bibliography{%
popsyn,popsyn_gw-merger-rates,popsyn-sed,%
observations-binaries-constraintsOnInteractions,%
star-formation-properties,%
observations-pulsars,observations-pulsars-kicks,observations-supernovae,%
gw-astronomy-mergers,gw-astronomy-bursts,%
mm-general,%
LIGO-publications,%
Astrophysics,astrophysics-stellar-dynamics-theory,short-grb,textbooks}

\end{document}